\documentclass{article}
\usepackage{frascatiphys}
\usepackage{graphicx}
\usepackage{boldline} 
\usepackage{xcolor} 
\usepackage{indentfirst}
\usepackage{hyperref}
\usepackage{cite}

\definecolor{darkgreen}{rgb}{0.0, 0.5, 0.13}
\definecolor{darkred}{rgb}{0.55, 0.0, 0.0}
\begin{document}
\title{
    TOWARDS HARDWARE ACCELERATION FOR PARTON DENSITIES ESTIMATION
}
\author{
    Stefano Carrazza \\
    Juan Cruz-Martinez \\
    Jes\'us Urtasun-Elizari \\
    Emilio Villa \\
    {\em TIF Lab, Dipartimento di Fisica, Universit\`a degli Studi di Milano and INFN Sezione di Milano,} \\
    {\em Via Celoria 16, 20133, Milano, Italy}
}
\maketitle
\baselineskip=11.6pt

\begin{abstract}
In this proceedings we describe the computational challenges associated to the determination of parton distribution functions (PDFs). We compare the performance of the convolution of the parton distributions with matrix elements using different hardware instructions. We quantify and identify the most promising data-model configurations to increase PDF fitting performance in adapting the current code frameworks to hardware accelerators such as graphics processing units.
\end{abstract}

\baselineskip=14pt

\section{Introduction}

The determination of parton distribution functions (PDFs) is a particular topic which strongly relies on three dynamic and time dependent factors: new experimental data, higher order theoretical predictions
and fitting methodology. In this environment, there are two main tasks for PDF fitters such as the NNPDF collaboration~\cite{AbdulKhalek:2019ihb,AbdulKhalek:2019bux,Ball:2018iqk}, MMHT~\cite{Harland-Lang:2014zoa} or CTEQ~\cite{Hou:2019jgw}. The first task consists in maintaining and organizing a workflow which incrementally implements new features proposed by the respective experimental and theoretical communities. The second task of a PDF fitter corresponds to investigate new numerical and efficient approaches to PDF fitting methodology. While the former relies almost exclusively on external groups and communities, the later is under full control of the PDF fitting collaborations, and in most cases it reflects the differences between them.

\begin{figure}
	\centering
	\includegraphics[scale=0.35]{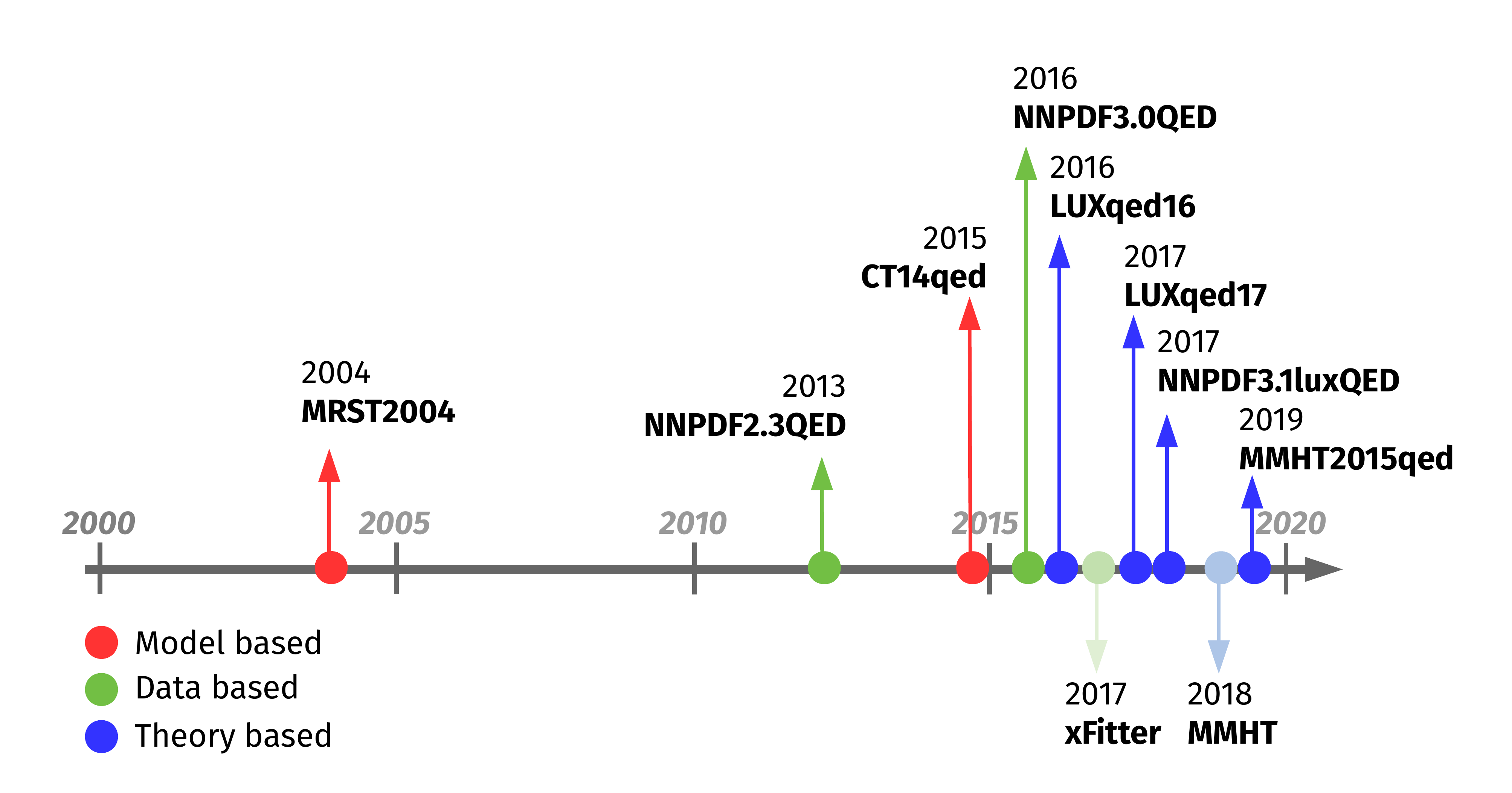}
	\caption{\label{timeline} Illustrative timeline of PDF releases with QED corrections since 2004. Arrows pointing up refer to publications providing PDF sets as deliverables.}
\end{figure}

As a real example of the previous description we show in figure~\ref{timeline} a non-exhaustive timeline for PDF determinations with QED corrections. Since 2004 we observe at least three different fitting approaches to the determination of the photon PDF, starting from model based approach where the photon PDF is modelled by an ad-hoc distribution~\cite{Martin:2004dh,Schmidt:2015zda}, then to a data driven approach where the photon PDF is extracted directly from data~\cite{Ball:2013hta,Bertone:2016ume,Giuli:2017oii}, and finally to a more precise procedure involving theory calculations~\cite{Manohar:2016nzj,Manohar:2017eqh,Bertone:2017bme,Nathvani:2018pys,Harland-Lang:2019pla}.
Improvements in terms of PDF quality and physics content are however accompanied by a negative performance trend associated to longer fitting times and increased computational cost due to larger datasets and increasingly complex procedures. This in turn makes more difficult the task of developing and testing novel fitting procedures.

In this proceedings we describe a new approach to deal with this growing trend in computational resources for PDF determination. We focus our discussion on the new methodology recently presented by the N3PDF team within the NNPDF collaboration and summarized in the next paragraphs.

\section{A deep learning approach to PDFs}

In Ref.~\cite{Carrazza:2019mzf} we presented a new approach to PDF fits based on deep learning techniques in the context of the NNPDF methodology. We implement a new efficient computing framework based on graph generated models for PDF parametrization and gradient descent optimization called {\tt n3fit}. The best model configuration is derived from a robust cross-validation mechanism through a hyperparametrization tune procedure. From a practical point of view the {\tt n3fit} code uses Keras~\cite{chollet2015keras} and TensorFlow (TF)~\cite{tensorflow2015:whitepaper} as backends.

\begin{figure}
	\centering
	\includegraphics[scale=0.52]{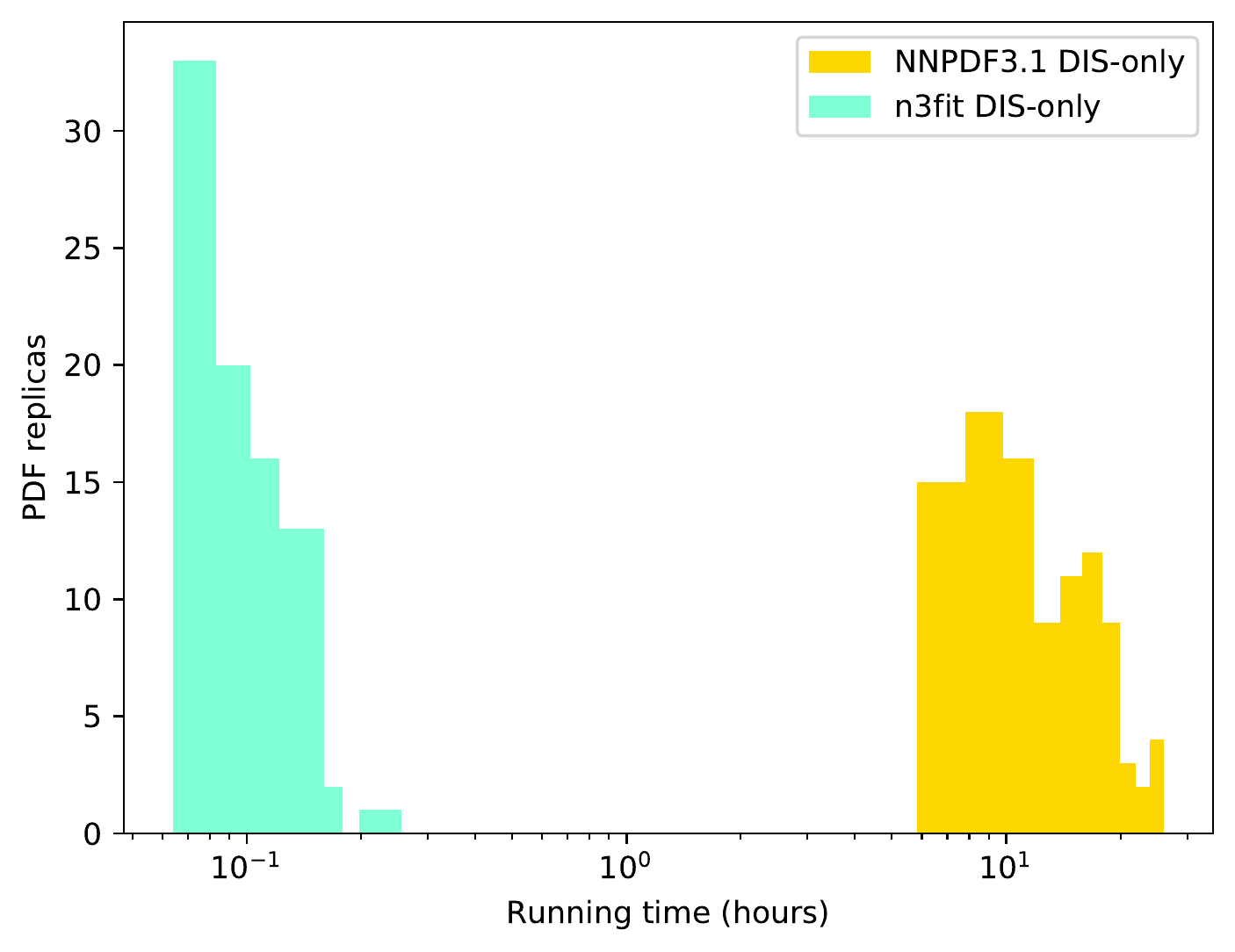}\includegraphics[scale=0.52]{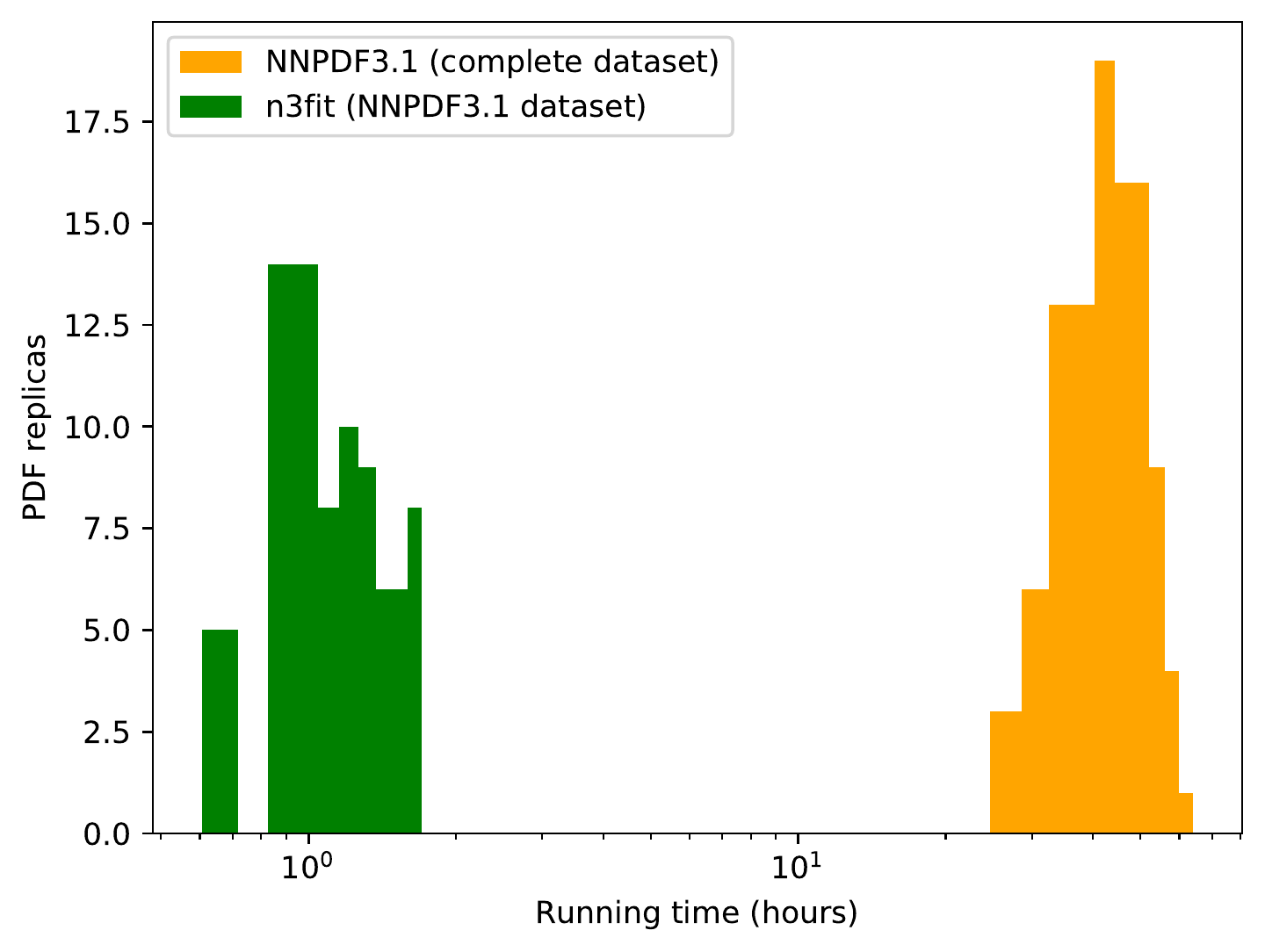}
	\caption{\label{times} Fitting time distribution per replica PDF for NNPDF and {\tt n3fit} codes for DIS-only (left plot) and global datasets (right plot).}
\end{figure}

From a technical perspective, one of the most relevant achievements of {\tt n3fit} is the reduction of computational time required to obtain PDF sets based on the NNPDF3.1 dataset~\cite{Ball:2017nwa}. In figure~\ref{times} we show the running time (in hours) required to fit 100 PDF replicas using the new {\tt n3fit} fitting code and we compare it to the latest {\tt NNPDF3.1} algorithm. On the left plot, we show a fit to DIS-only data, while in the right plot we have a global fit. In both cases we observe an improvement of between one to two orders of magnitude, {\it e.g.}~the new {\tt n3fit} code takes in average one hour to complete a global fit whereas the old code could take more than 40 hours. These improvements are partially due to the new minimizer (based on gradient descent instead of on genetic algorithms) in combination with multi-threading CPU calculations when executing the TensorFlow graph model.

The great performance improvement observed with {\tt n3fit} suggests that we may find new code strategies which take advantage of hardware acceleration. At this point, one may ask if it is possible improve the performance from the default TensorFlow graph optimization for CPU and eventually use hardware accelerators such as graphic processing units (GPUs), field-programmable gate arrays (FPGAs), or tensor processing units (TPUs).

\section{Hardware accelerating PDF determination}

The first step towards faster fits consists in profiling the code and isolate the most time consuming operations during PDF fits. Fortunately, the answer to this question is simple and involves the computation of physical observables through PDF and matrix elements convolutions,

\begin{equation}
	\sigma^{N} = \sum_{i, j, \alpha, \beta} f_{\alpha}(x_{i}) f_{\beta}(x_{j}) {\rm FK}^{N}_{ i j \alpha \beta} \;,
	\label{eq:observable}
\end{equation}
where $f_{\alpha}(x_{i})$ stands for the PDF of a particular flavor species $\alpha$ evaluated in the point $x_{i}$ of the grid in $x$. ${\rm FK}^{N}_{ i j \alpha \beta}$ is a Fast Kernel (FK) table which contains the information about the partonic cross section, following the description presented in~\cite{Carrazza:2019mzf}. In the case of hadronic observables the evaluation of predictions produces a vector of $N$ observables, $\sigma^{N}$, by building a neural network that generates the PDFs sampling from a grid in $x$, representing the fractions of momenta that a particular parton could carry, and then convoluting the result with an FK table containing the partonic information.

Given that TensorFlow relies on symbolic computation and graph generation to represent a model, as a first step we investigate if the memory usage it requires is higher than the one needed by a custom code specialized in convolutions.
We wrote a custom operator in {\tt C++} for TensorFlow that performs the convolution and its corresponding gradient, directly without graph evaluations. In table~\ref{table1} we cross-check the implementation by looking at examples of convolution and gradient computation for DIS and hadronic observables. The ratio between both implementations confirms excellent numerical agreement.
The memory usage for the default TensorFlow implementation and the custom convolution code are shown in table~\ref{table2}. We observe a reduction of 3.2 GB and 5.9 GB of resident memory when using our custom operator, when loading all NNPDF3.1 hadronic and global data respectively.
The reduction of memory usage is a great benefit because it gives the possibility to run PDF fits in consumer level hardware and, most importantly, load all data in the limited memory space available in hardware accelerators.
Once the memory saving is obtained, the performance can also be improved by multi-threading our custom operator on the CPU.

\begin{table}
  \centering
  \vspace{0.5cm}
	\begin{tabular}{|c|c|c|c|c|}
		\cline{3 - 5}
		\multicolumn{2}{c|}{ } & TensorFlow [pb] & Custom [pb] & Ratio \\
		\hline
		& & 1.9207904 & 1.9207904 & 1.0000000 \\
		DIS & Convolution & 2.4611666 & 2.4611664 &  0.9999999 \\
		& & 1.3516952 & 1.3516952 & 1.0000000 \\
		\cline{2 - 5}
		& & 1.8794115 & 1.8794115 & 1.0000000 \\
		& Gradient & 1.505316 & 1.505316 & 1.0000000 \\
		& & 2.866085 & 2.866085 & 1.0000000 \\
		\hline
		& & 8.142365 & 8.142366 & 1.0000001 \\
		Hadronic & Convolution & 8.947762 & 8.947762 & 1.0000000 \\
		& & 7.4513326 & 7.4513316 &  0.9999999 \\
		\cline{2 - 5}
		& & 18.525095 & 18.525095 & 1.0000000 \\
		& Gradient & 19.182995 & 19.182993 & 0.9999999 \\
		& & 19.551006 & 19.551004 & 0.9999999 \\
		\hline
	\end{tabular}
	\caption{\label{table1} Example of convolution and gradient computations for both DIS and hadronic observables. The ratio between TensorFlow and the custom computation confirms excellent numerical agreement.}
\end{table}

\begin{table}
	\centering
	\begin{tabular}{|c|c|c|c|c|}
		\cline{3 - 5}
		\multicolumn{2}{c|}{ } & TensorFlow  & Custom Convolution & Difference (TensorFlow - custom) \\
		\hline
		Hadronic & Virtual & {\color{darkred} 17.7 GB} & {\color{darkgreen} 13.8 GB} & {\color{darkgreen} 3.9 GB} \\
		& RES & {\color{darkred} 12.1 GB} & {\color{darkgreen} 8.39 GB} & {\color{darkgreen} 3.2 GB} \\ \hline
		Global & Virtual & {\color{darkred} 23.5 GB} & {\color{darkgreen} 19.7 GB} & {\color{darkgreen} 3.8 GB} \\
		& RES & {\color{darkred} 18.4 GB} & {\color{darkgreen} 12.5 GB} & {\color{darkgreen} 5.9 GB} \\ \hline
	\end{tabular}
	\caption{\label{table2} Memory usage after model generation and fit for both hadronic and global (DIS + hadronic).}
\end{table}

After the memory usage analysis, we carried out a time performance comparison running the convolution both on CPU and GPU. Shifting the computation from CPU to GPU one can take advantage of the parallelization over the increased number of cores. In table~\ref{table3} we show the overall running time for several examples of toy PDF and FK tables based on hadronic observables.
The numbers include the computation time as well as the time required for the memory transfer to the GPU.

The performance of Advanced Vector Extensions (AVX) on CPU, OpenCL~\cite{opencl} on GPU and TF both on CPU and GPU are compared. As it is shown in table~\ref{table3}, AVX and TF running on CPU are faster up to a certain number of columns (for the FK table) or rows (for the PDFs matrices). Once the size of the operation is big enough, AVX is over one order of magnitude slower than OpenCL and even two orders of magnitude slower than TF on GPU. TF on CPU is more resilient than our AVX implementation and it is competitive with the GPU convolutions for a larger range of dimensions.

According to these results, for given dimensions of the FK table and the PDF matrices, it is convenient to carry out the PDF fits on GPU devices, parallelizing, for instance, over the different PDF replicas, allowing the fit of all of them to run simultaneously.
In other words, hardware accelerators become competitive tools for PDF fitting once the penalty introduced by the memory transfer between devices is overcome.

\begin{table}
	\centering
	\renewcommand{\arraystretch}{1.2}
	\begin{tabular}{|m{2cm}|m{2.2cm}|m{2cm}|m{2cm}|m{2cm}|m{2cm}|}
		\hline
		Size of the PDF & Size of the FK Table &  TensorFlow CPU [s] & AVX [s] & TensorFlow GPU [s] & OpenCL [s] \\ \hline
		35721 $\times$ 1 & 8 $\times$ 35721 & 1.10 $\cdot 10^{-2}$ & \textcolor{darkgreen}{1.57 $\cdot 10^{-4}$} & 4.14 $\cdot 10^{-1}$ & $\sim$ 1 \\ \hlineB{2.5}
		$10^{6}$ $\times$ 1 & 8 $\times$ $10^{6}$ & 4.70 $\cdot 10^{-2}$ & \textcolor{darkgreen}{5.00 $\cdot 10^{-3}$} & 4.49 $\cdot 10^{-1}$ & $\sim$ 1 \\ \hline
		$10^{7}$ $\times$ 1 & 8 $\times$ $10^{7}$ & 3.20 $\cdot 10^{-1}$ & \textcolor{darkgreen}{5.70 $\cdot 10^{-2}$} & 7.90 $\cdot 10^{-1}$ & 1.38 \\ \hline
		$10^{8}$ $\times$ 1 & 8 $\times$ $10^{8}$ & 2.90 & \textcolor{darkgreen}{5.70 $\cdot 10^{-1}$} & 4.21 & 6.31 \\ \hlineB{2.5}
		35721 $\times$ $10^{2}$ & 8 $\times$ 35721 & 6.90 $\cdot 10^{-2}$ & \textcolor{darkgreen}{1.60 $\cdot 10^{-2}$} & 4.31 $\cdot 10^{-1}$ & $\sim$ 1 \\ \hline
		35721 $\times$ $10^{3}$ & 8 $\times$ 35721 & \textcolor{darkgreen}{1.50 $\cdot 10^{-1}$} & 1.69 $\cdot 10^{-1}$ & 5.63 $\cdot 10^{-1}$ & $\sim$ 1 \\ \hline
		35721 $\times$ $10^{4}$ & 8 $\times$ 35721 & \textcolor{darkgreen}{1.12} & 1.73 & 1.92 & 1.76 \\ \hline
		$35721 \times 5 \cdot 10^{4}$ & 10 $\times$ 35721 & \textcolor{darkgreen}{5.33} & 8.93 & 7.83 & 5.80 \\ \hlineB{2.5}
		35721 $\times$ 1 & $10^{2}$ $\times$ 35721 & 2.80 $\cdot 10^{-2}$ & \textcolor{darkgreen}{2.43 $\cdot 10^{-3}$} & 4.24 $\cdot 10^{-1}$ & $\sim$ 1 \\ \hline
		35721 $\times$ 1 & $10^{3}$ $\times$ 35721 & 1.30 $\cdot 10^{-1}$ & \textcolor{darkgreen}{2.14 $\cdot 10^{-2}$} & 5.60 $\cdot 10^{-1}$ & $\sim$ 1 \\ \hline
		35721 $\times$ 1 & $10^{4}$ $\times$ 35721 & 1.14 & \textcolor{darkgreen}{2.16 $\cdot 10^{-1}$} & 1.93 & 1.76 \\ \hlineB{2.5}
		35721 $\times$ $10^{2}$ & $10^{2}$ $\times$ 35721 & \textcolor{darkgreen}{6.20 $\cdot 10^{-2}$} & 1.86 $\cdot 10^{-1}$ & 4.32 $\cdot 10^{-1}$ & $\sim$ 1 \\ \hline
		35721 $\times$ $10^{3}$ & $10^{3}$ $\times$ 35721 & \textcolor{darkgreen}{3.00 $\cdot 10^{-1}$} & 21.61 & 7.19 $\cdot 10^{-1}$ & 5.25 \\ \hline
		$35721 \times 2 \cdot 10^{3}$ & $2 \cdot 10^{3} \times$ 35721 & 5.06 & 86.13 & \textcolor{darkgreen}{1.38} & 15.97 \\ \hlineB{2.5}
	\end{tabular}
  \caption{\label{table3} Time performances achieved with AVX, TensorFlow (both on CPU and GPU) and OpenCL for the given sizes of the FK table and the PDF matrix. In green is highlighted the lowest value within each row. Time is given in seconds. The base FK table used in this comparison consists on 49 flavour combinations and an grid in $x$ of size 27 (35721 elements in total).}
\end{table}

\section{Outlook and future developments}

The results presented in this proceedings strongly suggest that PDF fits can benefit from hardware accelerators such as GPUs, and in future, FPGAs or TPUs thanks to the possibility of offloading the most time-consuming tasks to the accelerator. However, we should notice that in order to achieve performance improvements some precautions are required by defining the sizes of FK tables and the number of PDFs we would like to convolute simultaneously.
In future work we are planning to extend the {\tt n3fit} framework to support GPU hardware.

\section*{Acknowledgements}

S.C. acknowledges the NVIDIA Corporation for the donation of a Titan V GPU used
for this research.
S.C., J.CM. and J.U. are supported by the European Research Council under the European
Unions Horizon 2020 research and innovation Programme (grant agreement number
740006). S.C. is supported by the UNIMI Linea 2A grant ``New hardware for HEP''.

\end{document}